\documentclass[12pt]{article}
\usepackage{authblk}
\usepackage{graphicx}
\usepackage{amssymb}
\usepackage{amsmath}
\title{Thermodynamics of AdS planar black holes and holography}

\author[a]{Neven Bili\'c\thanks{bilic@irb.hr}}
\author[b]{J\'ulio C.\ Fabris\thanks{julio.fabris@cosmo-ufes.org}}
\affil[a]{Division of Theoretical Physics, Rudjer Bo\v skovi\'c Institute, 10002 Zagreb, Croatia}
\affil[b]{N\'ucleo Cosmo-ufes \& Departamento de F\'isica, Universidade Federal do Esp\'irito Santo (UFES)
	Av.\ Fernando Ferrari s/n CEP 29.075-910, Vit\'oria, ES, Brazil}

\date{\today}
\begin{document}
\maketitle
\begin{abstract}

A large AdS  Schwarzschild black hole can be approximated by an AdS planar black hole.
However, the temperature of a planar black hole is not well defined due to the translational  invariance of the horizon.
 We propose to fix this arbitrariness by imposing  
the entropy area law. 
Furthermore, using the AdS/CFT holography, we propose a relationship between the 
temperatures of an AdS planar and large AdS Schwarzschild  
black holes.

%Using the holographic AdS/CFT prescription we derive the temperature of the large AdS Shwazshild black hole.

\end{abstract}

%\pacs{03.75.Kk, 04.40.Nr, 47.10.-g, 98.80.Jk}
%
%\keywords{analog gravity, cosmology, fluid dynamics, Bose-Einstein condensate}

%\maketitle

\section{Introduction}
%One of the essential ingridients of General Relativity is the  geometry of spacetime.

%In this paper we extend these ideas to a more general case that includes nonisentropic fluids.

Anti-de Sitter (AdS) Schwarzschild black holes have a curious thermodynamic property. 
The horizon temperature of a large AdS Schwarzschild black hole (BH) increases with increasing horizon radius in contrast to an ordinary Schwarzschild BH 
the temperature of which decreases with horizon radius. Hubeny, Marolf, and  Rangamani  \cite{hubeny} argue that the thermodynamic behavior 
of AdS Schwarzschild BHs is tightly related to the thermodynamic properties of planar BHs. This argument is based on the fact  
that a large AdS Schwarzschild BH can be well approximated by an AdS planar BH \cite{witten2}.

 To the best of our knowledge, the AdS planar BHs were first stadied by Lemos
\cite{lemos1} and more recently  were shown to appear
by scaling  a near-extremal brane solution \cite{horowitz2}. 
Interesting properties of AdS planar BHs  were found \cite{lemos2} in the context of  thermodynamics.  
Besides, 
planar BHs  
have recently attracted considerable attention in non-gravitational contexts. 
Geometric structures in the form of a planar BH may have interesting
applications 
in condensed matter physics \cite{hartnoll},
2+1-dimensional superconductor
\cite{bobev,albash,chakraborty,bilic4}, and acoustic geometry \cite{bilic4,hossenfelder2,nikolic}.

As we will shortly demonstrate,  
the horizon of an AdS planar BH is translated when a planar BH metric
undergoes a rescaling transformation, so the horizon location of an AdS planar BH is not well defined.
Following Hubeny, Marolf, and Rangmani \cite{hubeny} we will refer to this property as "translational invariance" of the BH horizon, even though  
this term could be somewhat misleading in this context.   Based on this property, the authors of Ref.\ \cite{hubeny} 
argue that large Schwarzschild-AdS black holes, although very hot,
locally appear at most only lukewarm.

In this paper, we study the relationship between a large AdS Schwarzschild BH in
$d+1$ dimensions
and  AdS planar BH in the context of anti-de Sitter/conformal field theory (AdS/CFT) holography.
 First, we will show that the horizon temperature of a planar BH can be
	fixed by the requirement of the entropy area law. 
	Then, by invoking the AdS/CFT correspondence conjecture, we will relate the 
	local temperatures of the planar BH and large AdS Schwarzschild BH near the AdS boundary.

We divide the remainder of the paper into three sections.
In section \ref{adsschw} we briefly review the thermal properties of AdS Schwarzschild and planar BHs.
In section \ref{holography} we derive relationships between the temperatures of the boundary conformal fluid and of a 
large AdS BHs by employing  AdS/CFT holography.
Concluding remarks are given in section \ref{conclude}.
%Finally, in appendix \ref{second} we provide a brief account of the second law of thermodynamics
%relevant for nonisentropic fluid flows.

\section{AdS Schwarzschild and planar black hole}
\label{adsschw}

The metric of an AdS Schwarzschild black hole in $d+1$ dimensions is usually written in the form
\cite{witten2,horowitz}
\begin{eqnarray}
ds^2 = f(r)  dt^2 - \frac{1}{f(r)} dr^2 -r^2 d\Omega_{d-1}.
\label{eq200}
\end{eqnarray}
The function $f$ is defined as
\begin{equation}
f(r)=\frac{r^2}{\ell^2}+1-\mu\left(\frac{\ell}{r}\right)^{d-2},
\label{eq1001}
\end{equation}
where $\ell$ is the curvature radius of AdS$_{d+1}$ and the dimensionless parameter $\mu$
is  related to the black-hole mass via \cite{witten2,myers}
\begin{equation}
\mu=\frac{16\pi G_{d+1} M_{\rm bh}}{(d-1)\ell^{d-2}\Omega_{d-1}}.
 \label{eq3105}
\end{equation}
Here, $G_{d+1}$ is the $d + 1$-dimensional Newton's constant and
\begin{equation}
\Omega_{d-1}=\frac{2\pi^{d/2}}{\Gamma(d/2)} =\left\{ \begin{array}{ll}
(2\pi)^{d/2}/(d-2)!!,
& \mbox{ for even $d$},\\
2(2\pi)^{(d-1)/2}/(d-2)!! ,
& \mbox{ for odd $d$}, \end{array} \right.
\label{eq4105}
\end{equation}
 is the volume of a
unit $d - 1$-sphere.
The parameter $\mu$ can be expressed in terms of  the largest root $r_+$ of the equation $f(r)=0$,
\begin{equation}
\mu=\left(\frac{r_+}{\ell}\right)^{d-2}\left(1+\frac{r_+^2}{\ell^2}\right).
\label{eq1002}
\end{equation}
Note that the metric (\ref{eq200}) with (\ref{eq1001}) with  horizon radius involves two scales: the AdS curvature $\ell$ and
 $d+1$-dimensional Newton's constant $G_{d+1}$

%\subsection{Horizon temperature}
The horizon temperature af an AdS Schwarzschild BH can be derived in the usual way by
approximating the Euclidean metric in the near horizon limit with
\begin{equation}
f=(r-r_+)f_+ ,
\end{equation}
where
\begin{equation}
f_+=\left. \frac{\partial f}{\partial r}\right|_{r=r_+}.
\end{equation}
Defining the periodicity of the Euclidean time so to avoid the
cone singularity, one finds the inverse horizon temperature
\begin{equation}
\beta=\frac{1}{T_+}=\frac{4\pi}{f_+}=
\frac{4\pi \ell^2 r_+}{d r_+^2+(d-2)\ell^2}.
\label{eq1010}
\end{equation}
Note that,  the horizon temperature $T_+\rightarrow \infty$ when either $r_+\rightarrow 0$
(small BH)  or
$r_+\rightarrow \infty$ (large BH) unlike in the Schwarzschild BH in an asymptotically flat background where $T_+\rightarrow 0$
for a large BH.
In particular, for a large AdS Schwarzschild BH we have 
\begin{equation}
T_+=\frac{d r_+}{4\pi \ell^2}.
\label{eq2010}
\end{equation}

\subsection{AdS Planar black hole}
As pointed by Witten \cite{witten2}, in the limit of large AdS Schwarzschild BH the topology goes from
${\bf S}^1\times {\bf S}^{d-1}$ to ${\bf S}^1\times {\bf R}^{d-1}$ where a Schwarzschild BH is approximated by a planar BH
with a translationally  invariant horizon.
To see this, consider a large Schwarzschild-AdS black hole with $\mu\gg 1$ ($r_+\gg \ell$).
By rescaling
\begin{equation}
r=\mu^{1/d} \frac{\ell^2}{\rho},
\label{eq1003}
\end{equation}
the function $f$ becomes
\begin{equation}
f= \mu^{2/d}\left(\frac{\ell^2}{\rho^2}-\frac{\rho^{d-2}}{\ell^{d-2}}+\mu^{-2/d}\right).
\label{eq1004}
\end{equation}
Then, by neglecting the last term in  brackets and  rescaling the time $t=\mu^{-1/d}\tau$,
the line element squared becomes
\begin{equation}
ds^2 =\frac{\ell^2}{\rho^2}\left[\left(1-\frac{\rho^d}{\ell^d}\right)d\tau^2
-\left(1-\frac{\rho^d}{\ell^d}\right)^{-1} d\rho^2
-\mu^{2/d}\ell^2 d\Omega_{d-1}^2\right] .
\label{eq1005}
\end{equation}
For $M_{\rm bh} \rightarrow \infty$ (i.e., for $\mu \rightarrow \infty$) the radius $\mu^{1/d}\ell$ of
${\bf S}^{d-1}$
 diverges and hence, ${\bf S}^{d-1}$ becomes flat and looks locally as ${\bf R}^{d-1}$.
 Now we introduce coordinates $x_i$ near a point $P\in  {\bf S}^{d-1}$  such that at $P$,
 $d\Omega_{d-1}^2=\ell^{-2}\mu^{-2/d}d\mbox{\boldmath $x$}^2$. Then the metric becomes
 \begin{equation}
ds^2 =\frac{\ell^2}{\rho^2}\left[\left(1-\frac{\rho^d}{\ell^d}\right)d\tau^2
-\left(1-\frac{\rho^d}{\ell^d}\right)^{-1} d\rho^2
-d \mbox{\boldmath $x$}^2\right] ,
\label{eq1006}
\end{equation}
where
\begin{equation}
 d \mbox{\boldmath $x$}^2=  \sum_{i=1}^{d-1} {\rm d} x^i {\rm d} x^i .
\end{equation}
 The metric (\ref{eq1006}) describes an asymptotically AdS$_{d+1}$ geometry with a planar BH
horizon located at $\rho=\ell$.
Note that dependence on Newton's constant $G_{d+1}$ has disappeared and the metric  (\ref{eq1006}) involves only one scale: the AdS curvature $\ell$.

Now, it is easy to show  by simultaneously rescaling the coordinates
that the metric (\ref{eq1006}) is equivalent to the whole class of 
planar BHs represented by the line element 
\begin{equation}
ds^2 =\frac{\ell^2}{\rho^2}\left[\left(1-\frac{\rho^d}{\rho_{\rm pl}^d}\right)d\tau^2
-\left(1-\frac{\rho^d}{\rho_{\rm pl}^d}\right)^{-1} d\rho^2
-d \mbox{\boldmath $x$}^2\right] ,
\label{eq1007}
\end{equation}
with horizon located at an arbitrary $\rho_{\rm pl}$.

As before, the horizon temperature of the Planar BH can be calculated
by demanding that the periodicity $\beta_{\rm pl}$ is such that
the conical singularity is removed.
One finds
\begin{equation}
	\beta_{\rm pl}=\frac{1}{T_{\rm pl}}=\frac{4\pi\rho_{\rm pl}}{d},
	\label{eq1011}
\end{equation}
so the temperature of the planar BH increases when the horizon shifts
towards the AdS boundary at $\rho=0$.
However, since $\rho_{\rm pl}$ is not fixed, 	the temperature $T_{\rm pl}$  is not well defined. Hence, local observations are independent of $\rho_{\rm pl}$ and, since the AdS scale is the only scale in the problem, local observers see no excitations above this scale.
	
	The fact that the temperature $T_{\rm pl}$ of the AdS planar BH is not well defined, is reminiscent of the case for the extreme Reissner-Nordstr\"om BH
	with AdS near horizon geometry. 
	As argued, e.g., in Refs.\ \cite{hawking2,fabris}, 	
	the temperature of the extreme Reissner-Nordstr\" om BH is arbitrary since its 
	horizon is infinitely far away.

Since small $\rho$ corresponds to large $r$, one would naively conclude that
the thermodynamic behavior of the AdS planar BH is consistent with that of the
large AdS Schwarzschild BH. However, this conclusion could be
wrong as it is not obvious
that the planar BH temperature is in any way related to the temperature of the AdS Schwarzschild BH.
%due to arbitrariness of $\rho_{\rm pl}$.

\subsection{Entropy of AdS planar black holes}
\label{entropy}

We closely follow Witten's procedure   for
the calculation of the BH
entropy. 
This procedure applied to the metric 
(\ref{eq200}) with (\ref{eq1001}) yields the AdS Schwarzschild BH entropy \cite{witten2} 
\begin{equation}
	S_{\rm BH}= \frac{r_+^{d-1} \Omega_{d-1}}{4 G_{d+1}},
	\label{eq1046}
\end{equation}
which demonstrates the area law in agreement with Bekenstein \cite{bekenstein}
and Hawking \cite{hawking1}.
To calculate the horizon entropy of the  AdS planar BH 
we start from the metric (\ref{eq1007})
in which we transform the coordinate $\rho$ to $r=\rho_{\rm pl}^2/\rho$ so that $r$ goes from $r=\rho_{\rm pl}$
to $r=\infty$ as $\rho$ goes from $\rho=\rho_{\rm pl}$ to $\rho=0$.
The Euclidean metric is given by
\begin{equation}
ds_{\rm E}^2 =\frac{r^2}{\ell^2}\left[\left(1-
\frac{\rho_{\rm pl}^d}{r^d}\right)d\tau^2
+\left(1-
\frac{\rho_{\rm pl}^d}{r^d}\right)^{-1} \frac{\ell^4}{r^4}dr^2
+d \mbox{\boldmath $x$}^2\right] ,
\label{eq2001}
\end{equation}
with the associated horizon temperature $T_{\rm pl}$ given by (\ref{eq1011}).

Following Hawking and Page \cite{hawking} and Witten \cite{witten2}
we describe the BH free energy as a renormalized on shell Euclidean  bulk action.
The on shell AdS BH action is proportional to an infinite volume of the AdS BH spacetime
so it must be regularized and renormalized.
The regularization is achieved by integrating the space volume up to a
large radius $R$.
The renormalized action is then obtained by subtracting the pure AdS space time volume from the
AdS BH spacetime volume, assuming equality of the local AdS temperature
and the redshifted planar BH temperature at the hypersurface  $r=R$.
More explicitly
\begin{equation}
F_{\rm pl}=-T_{\rm pl} \ln Z=\frac{d\ell^{-2}}{8\pi G_{d+1}\beta_{\rm pl}}\lim_{R\rightarrow\infty}\left(V_{\rm pl}-V_{\rm AdS}\right) ,
\label{eq1040}
\end{equation}
where
\begin{equation}
V_{\rm pl}=\int_0^{\beta_{\rm pl}} d\tau\int_0^R dr \sqrt{g} \left(\int_{-L/2}^{L/2} dx\right)^{d-1} ,
\label{eq1041}
\end{equation}
\begin{equation}
V_{\rm AdS}=\int_0^{\beta_{\rm AdS}} d\tau\int_0^R dr \sqrt{g} \left(\int_{-L/2}^{L/2} dx\right)^{d-1} .
\label{eq1042}
\end{equation}
The length $L$ is the linear size of the hyperplane $r={\rm const.}$.
The inverse AdS temperature $\beta_{\rm AdS}$ is obtained from the equation
\begin{equation}
\beta_{\rm AdS}\sqrt{g_{00}^{\rm AdS}}=\beta_{\rm pl}\sqrt{g_{00}^{\rm pl}} ,
\label{eq1043}
\end{equation}
where $g_{00}^{\rm AdS}$ and $g_{00}^{\rm pl}$ are the time-time components of the AdS and-black hole
metric tensors, respectively, evaluated at $r=R$. We find
\begin{equation}
\beta_{\rm AdS}=\beta_{\rm pl}\left(1-\frac{\rho_{\rm pl}^d}{R^d}\right)^{1/2} .
\label{eq1044}
\end{equation}
Then, from(\ref{eq1040}) we obtain the free energy expressed as a function of $\rho_{\rm pl}$
\begin{equation}
F_{\rm pl}=\frac{L^{d-1}}{16\pi G_{d+1}\ell}\left(\frac{\rho_{\rm pl}}{\ell}\right)^d ,
\label{eq1045}
\end{equation}
where $L^{d-1}$ is the area of the hyperplane $r= {\rm const.}$.
From this, using (\ref{eq1011}) and the thermodynamic relation between entropy and free energy,
we find
\begin{equation}
S_{\rm pl}= -\frac{\partial F_{\rm pl}}{\partial T_{\rm pl}}=\frac{L^{d-1}}{4 G_{d+1}}
\left(\frac{\rho_{\rm pl}}{\ell}\right)^{d+1} .
\label{eq3046}
\end{equation}
Hence,  we recover the standard area law for the planar BH entropy if $\rho_{\rm pl}=\ell$.
So, if we demand the validity of the area law, the metric (\ref{eq1005}) will be a natural
representative of all AdS planar BHs belonging to the class described by
the line element (\ref{eq1006}) with $0<\rho_{\rm pl}<\infty$.

We shall next invoke the holographic principle to relate the temperature 
$T_{\rm pl}$ of the planar BH  to the
temperature of the conformal fluid at the boundary of AdS spacetime.

\section{AdS/CFT holography}
\label{holography}

AdS/CFT holographic principle relates the bulk geometry to the conformal field theory at the AdS boundary.
To study the CFT at the boundary, it is convenient to transform both the AdS Schwarzschild metric (\ref{eq200})
and the AdS planar BH metric (\ref{eq1007}) to the Fefferman-Graham coordinates.
In Fefferman-Graham coordinates
the general asymptotically AdS metric can be expressed as 
\begin{equation}
	ds^2=\frac{\ell^2}{z^2}\left( g_{\mu\nu}(z) dx^\mu dx^\nu -dz^2\right),
	\label{eq3001}
\end{equation}
where the $d$-dimensional metric  $g_{\mu\nu}$  near the boundary at $z=0$ can be expanded as \cite{haro}
\begin{equation}
	g_{\mu\nu}=g^{(0)}_{\mu\nu}+z^2 g^{(2)}_{\mu\nu}
	+\ldots \, . +z^d g^{(d)}_{\mu\nu} +h_{\mu\nu}  z^d \ln z^2 +\mathcal{O}(z^{d+1}).
	\label{eq3002}
\end{equation}
The logarithmic term appears only for $d$ even.
Using this expansion one can express  
the vacuum expectation value of the boundary CFT stress tensor as 
\cite{haro}
\begin{equation}
	\langle T_{\mu\nu}\rangle = - \frac{d\ell^{d-1}}{16\pi G_{d+1}}g^{(d)}_{\mu\nu}+
	X_{\mu\nu}[g^{(n)}] ,
	\label{eq1017}
\end{equation}
where $X_{\mu\nu}[g^{(n)}]$ is a function of $g^{(n)}_{\mu\nu}$ with $n < d$. Its exact form depends on the spacetime
dimension and it reflects the conformal anomalies of the boundary CFT \cite{henningson}
(for $d=4$ and applications in cosmology see also Refs.\ \cite{apostolopoulos,bilic1}).
For  $d$ odd, there are no gravitational conformal anomalies
and $X_{\mu\nu}$ is equal to zero. 
For $d$ even, we may expect that the conformal anomaly related to the $X$ term will correspond to the logarithmic correction to the BH entropy
\cite{fursaev,cai,chang}.
In the following, we will be interested in the dominant area law term in the BH entropy for which we expect to correspond  to
the traceless part of the universal term in (\ref{eq1017}) proportional to $g^{(d)}_{\mu\nu}$.

In the following, we detail the correspondence between the vacuum expectation value of the CFT stress tensor 
and the geometries of Schwarzschild and planar black holes via equation (\ref{eq1017}).

\subsection{AdS Schwarzschild BH}
\label{schwarzschild}
Consider first the AdS Schwarzschild BH.
 By transforming the radial coordinate $r=\ell^2/\rho$, we rewrite the metric
(\ref{eq200}) in the form similar to (\ref{eq1005}), 
\begin{equation}
	ds^2 =\frac{\ell^2}{\rho^2}\left[\left(1+\frac{\rho^2}{\ell^2}-\mu\frac{\rho^d}{\ell^d}\right)d\tau^2
	-\left(1+\frac{\rho^2}{\ell^2}-\mu\frac{\rho^d}{\ell^d}\right)^{-1} d\rho^2
	-\ell^2 d\Omega_{d-1}^2\right] .
	\label{eq1022}
\end{equation}
Then, the transformation
\begin{equation}
	\frac{dz}{z}=\frac{d\rho}{\rho}\left(1+\frac{\rho^2}{\ell^2}-\mu \frac{\rho^d}{\ell^d}\right)^{-1/2}
	\label{eq1023}
\end{equation}
will take us to the Fefferman-Graham form. 
The integration of (\ref{eq1023}) cannot be performed in terms of elementary functions, with the exception of $d=2$ and $d=4$ \cite{apostolopoulos}.
However, as we are interested in the expansion (\ref{eq3002}) near $z=0$ up to the order $z^d$, we can
apply an ansatz to the metric in Fefferman-Graham coordinates:
\begin{equation}
	ds^2 =\frac{\ell^2}{z^2}\left[H(z)d\tau^2
	-F(z)\ell^2 d\Omega_{d-1}^2 -dz^2\right] .
	\label{eq1024}
\end{equation}
Here,
\begin{equation}
	F(z)=1+\sum_{n=2}^{\infty} f_n \frac{z^n}{\ell^n},
	\quad
	H(z)=1+\sum_{n=2}^{\infty} h_n \frac{z^n}{\ell^n},
	\label{eq1025}
\end{equation}
where the coefficients $f_n$ and $h_n$ are to be determined by
equating the line element (\ref{eq1024}) with (\ref{eq1022}) and
making use of (\ref{eq1023}). We find
\begin{equation}
	F(z)=1-\frac12 \frac{z^2}{\ell^2}+\frac{1}{16}\frac{z^4}{\ell^4}
	+\frac{1}{d}\mu \frac{z^d}{\ell^d}+\dots ,
	\label{eq1026}
\end{equation}
\begin{equation}
	H(z)=1+\frac12 \frac{z^2}{\ell^2}+\frac{1}{16}\frac{z^4}{\ell^4}
	-\frac{d-1}{d}\mu \frac{z^d}{\ell^d}+\dots ,
	\label{eq1027}
\end{equation}
so the background is the static Einstein universe ${\bf R}\times {\bf S}^{d-1}$ with line element
\begin{equation}
	g^{(0)}_{\mu\nu}dx^\mu dx^\nu=d\tau^2- \ell^2d\Omega_{d-1}^2
	% g^{(d)}_{\mu\nu}dx^\mu dx^\nu=-\frac{\mu}{d\ell^d}[(d-1)d\tau^2+ \ell^2d\Omega_{d-1}^2],
	\label{eq1028}
\end{equation}

For $d\neq 4$ we find
\begin{equation}
	% g^{(0)}_{\mu\nu}dx^\mu dx^\nu=d\tau^2- \ell^2d\Omega_{d-1}^2, \quad\quad
	g^{(d)}_{\mu\nu}dx^\mu dx^\nu=-\frac{\mu}{d\ell^d}[(d-1)d\tau^2+ \ell^2d\Omega_{d-1}^2],
	\label{eq1029}
\end{equation}
with trace ${\rm Tr} g^{(d)}=0$.
For $d=4$ we have
\begin{equation}
	g^{(4)}_{\mu\nu}dx^\mu dx^\nu=-\frac{1}{4\ell^4}
	\left[\left(3\mu-\frac{1}{4}\right)d\tau^2+
	\left(\mu+\frac{1}{4}\right)\ell^2d\Omega_{3}^2\right],
	\label{eq1030}
\end{equation}
with  ${\rm Tr} g^{(4)}=1/(4\ell^4)$.

To find the traceless part of the holographic stress tensor (\ref{eq1017}) we just subtract the trace, so for any $d\geq 2$ we have
\begin{equation}
	\langle T^{(d)}_{\mu\nu}\rangle = - \frac{d\ell^{d-1}}{16\pi G_{d+1}}\left(g^{(d)}_{\mu\nu}
	-\frac{1}{d} {\rm Tr} g^{(d)}g^{(0)}_{\mu\nu}\right).
	\label{eq1031}
\end{equation}
By writing  the BH mass  parameter $\mu$ in terms of $r_+$  by Eq.\  (\ref{eq1002}), 
we obtain
\begin{equation}
	\langle {{T^{(d)}}^\mu}_\nu\rangle = \frac{\ell^{d-1}}{16\pi G_{d+1}}
	\left(\frac{r_+}{\ell^2}\right)^d \left(1+\frac{\ell^2}{r_+^2}\right)
	\;{\rm diag} \left( d-1, -1,-1, ... \right).
	\label{eq1032}
\end{equation}
This stress tensor is of the form of the stress tensor of a $d$-dimensional
conformal fluid (CF) with the equation of state
\begin{equation}
	p_{\rm CF}=\varepsilon_{\rm CF}/(d-1) ,
\end{equation}
where $p_{\rm CF}$ and $\varepsilon_{\rm CF}$ are the fluid pressure and energy density, respectively, and
\begin{equation}
	\varepsilon_{\rm CF}= \frac{(d-1)\ell^{d-1}}{16\pi G_{d+1}}
	\left(\frac{r_+}{\ell^2}\right)^d\left(1+\frac{\ell^2}{r_+^2}\right).
	\label{eq1035}
\end{equation}

We now assume that the conformal fluid at the boundary is equivalent to
the black-body radiation at a temperature $T$. For a theory
with  $n_{\rm B}$ massless  bosons and $n_{\rm F}$  massless  fermions in $d-1$ spatial dimensions,  
the energy  density of a gas at a temperature $T$ is given by 
\begin{equation}
	\varepsilon= 
	\int \frac{d^{d-1}k}{(2\pi)^{d-1}} \left(\frac{n_{\rm B} k}{e^{k/T}-1}
+	 \frac{g n_{\rm F}k}{e^{k/T}+1}\right)
	=a_d T^d ,
	\label{eq1021}
\end{equation}
the pressure by $p=\varepsilon/(d-1)$,
and the entropy density by
\begin{equation}
	s = T^{-1} ( p+\varepsilon )
	=\frac{da_d}{d-1} T^{d-1} , 
	\label{eq1121}
\end{equation}
where  
\begin{equation}
	a_d=\frac{\Omega_{d-2}\Gamma(d) \zeta(d)}{(2\pi)^{d-1}}
\left( n_{\rm B} +  (1-2^{1-d})gn_{\rm F}\right)	
	.
	\label{eq2021}
\end{equation}
The factor $g$ is included to account for 
degeneracy due to internal degrees of freedom of the
fermions. For example, for Weyl fermions in $4$-dimensional spacetime,
$g=2$.

For a large BH ($r_+\gg \ell$), the energy density 
$\varepsilon_{\rm CF}$  (as expressed in Eq.\ (\ref{eq1035})) takes the form (\ref{eq1021})  with 
 the black-body temperature $T\propto r_+$. Then, according to (\ref{eq2010}),
the temperature $T$ of the conformal fluid at the AdS boundary 
is proportional to the horizon temperature $T_+$ 
of the large AdS Schwarzschild BH. 
%  Hence we can relate the BH temperature $T_+$ to the black-body temperature $T$ 
Hence,  the entropy density (\ref{eq1121}) of the conformal fluid at the boundary scales as $T_+^{d-1}$ as expected \cite{witten2}.

\subsection{AdS Planar BH}
\label{planar}
Next, we apply the above procedure to a planar BH.
 Starting from (\ref{eq1007}) we transform the coordinate $\rho$
into $z$
by
\begin{equation}
\frac{dz}{z}=\frac{d\rho}{\rho}
\left(1-\frac{\rho^d}{\rho_{\rm pl}^d}\right)^{-1/2} .
\label{1012}
\end{equation}
Integrating this equation we find
\begin{equation}
\left(\frac{z}{z_{\rm pl}}\right)^d= \frac{1-\sqrt{1-(\rho/\rho_{\rm pl})^d}}{1+\sqrt{1-(\rho/\rho_{\rm pl})^d}} ,
\label{1013}
\end{equation}
where $z_{\rm pl}$ is an arbitrary positive integration constant
which can be interpreted as the location of the planar BH horizon in Fefferman-Graham coordinates.
In the limit $\rho \rightarrow 0$ we have
\begin{equation}
(z/z_{\rm pl})^d=\frac14 (\rho/\rho_{\rm pl})^d +\mathcal{O}((\rho/\rho_{\rm pl})^{2d}) ,
\label{eq1014}
\end{equation}
 so it is natural to identify $z_{\rm pl}^d\equiv 4 \rho_{\rm pl}^d$.
 The inverse to (\ref{1013}) is easily found:
 \begin{equation}
\rho= z\left(
1+ \frac14 \left(\frac{z}{\rho_{\rm pl}}\right)^d \right)^{-2/d}.
\label{eq1015}
\end{equation}
 Using this we can recast the metric (\ref{eq1007}) in the Fefferman-Graham form
 \begin{equation}
ds^2 =\frac{\ell^2}{z^2}\left[\left(1-\frac14 \frac{z^d}{\rho_{\rm pl}^d}\right)^2
\left(1+\frac14 \frac{z^d}{\rho_{\rm pl}^d}\right)^{4/d-2}d\tau^2
-\left(1+\frac14 \frac{z^d}{\rho_{\rm pl}^d}\right)^{4/d}d \mbox{\boldmath $x$}^2
- dz^2\right] ,
\label{eq1016}
\end{equation}
We now compare this with the general expression for the asymptotically AdS metric
(\ref{eq3001}).
Expanding the functions in (\ref{eq1016}) and comparing with (\ref{eq3001}) and (\ref{eq3002})
we find
\begin{equation}
 g^{(0)}_{\mu\nu}=\eta_{\mu\nu}, \quad\quad g^{(d)}_{\mu\nu} = -\frac{1}{d\rho_{\rm pl}^d}\;{\rm diag} \left( d-1, 1, 1, ... \right) ,
\label{eq1018}
\end{equation}
so the traceless part of the stress tensor is given by
 \begin{equation}
 \langle {{T^{(d)}}^\mu}_\nu\rangle = \frac{\ell^{d-1}\rho_{\rm pl}^{-d}}{16\pi G_{d+1}}\;{\rm diag} \left( d-1, -1,-1, ... \right).
\label{eq1019}
\end{equation}
As in the case of the AdS Schwarzschild BH, 
this form is identical to the stress tensor of the $d$-dimensional
conformal fluid with the equation of state
$p_{\rm CF}=\varepsilon_{\rm CF}/(d-1)$, where the energy density is given by
\begin{equation}
 \varepsilon_{\rm CF}= \langle {{T^{(d)}}^0}_0\rangle = \frac{(d-1)\ell^{d-1}\rho_{\rm pl}^{-d}}{16\pi G_{d+1}}.
 \label{eq1020}
\end{equation}
We can compare this with  the energy density (\ref{eq1021}) of black-body radiation and relate $\rho_{\rm pl}$ to the black-body temperature $T\propto \rho_{\rm pl}^{-1}$. As before, we find that the entropy of the conformal fluid at the boundary scales as $T_{\rm pl}^{d-1}$.

\subsection{Temperature relationships}

In this section,  based on the correspondence between the CFT entropy at the boundary and the BH entropies, 
 we establish relationships between the temperatures $T$, $T_+$, and $T_{\rm pl}$, of the boundary conformal fluid, the 
large AdS Schwarzschild BH, and AdS planar BH, respectively.

	Consider first a large AdS Schwarzschild BH.
Our approach relies on the assumption that
the entropy of a large BH can be identified with the entropy of the conformal boundary. This assumption is quite natural since we have explicitly demonstrated that  the energy density of the conformal fluid at the boundary, according to equation (\ref{eq1035}), 
 scales as 
$r_+^d$   or equivalently as $T_+^d$. As a consequence, the entropy density of the conformal fluid  at the boundary scales as $T_+^{d-1}$, exactly as 
the  entropy of a large AdS Schwarzschild BH.
Hence, we equate the Schwarzschild  BH entropy $S_{\rm BH}$ given by (\ref{eq1046}) with the total entropy of the black-body radiation  on the boundary.
\begin{equation}
	S= \lim_{R\rightarrow \infty} \frac{d a_d}{d-1}T^{d-1} R^{d-1} \Omega_{d-1} ,
	\label{eq2046}
\end{equation}
where $R^{d-1}\Omega_{d-1}$ is the area of the large hypersphere of radius $R$, which approaches the boundary.
From the equation 
$S=S_{\rm BH}$ we find  
\begin{equation}
	T^{d-1}=\frac{d-1}{4d a_d G_{d+1}}	\frac{r_+^{d-1}}{R^{d-1}} . 
	\label{eq2047}
\end{equation}
Then, for a large BH ($r_+\gg \ell$), using (\ref{eq2010}) we find a relationship 
between the temperatures $T$ and $T_+$
\begin{equation}
	T=\frac{4\pi \ell}{d}\left( \frac{d-1}{4d a_d G_{d+1}}\right)^{1/(d-1)}    	\frac{T_+}{R/\ell} . 
	\label{eq2048}
\end{equation}
Note that, according to Tolman's relation applied to the asymptotic AdS spacetime  
\cite{page}, the local temperature near the boundary is given by
\begin{equation}
	T_{+,\rm loc} \equiv  	\frac{T_+}{\sqrt{g_{00}}}= \frac{T_+}{R/\ell}, 
	\label{eq2049}
\end{equation}
and hence, the expression (\ref{eq2048}) relates the temperature of the holographic conformal fluid to the local Schwarzschild BH temperature  near the  AdS boundary at $r=R$. 

Consider next the planar BH geometry. In this case,
 the total entropy of the black-body radiation  on the boundary is given by
\begin{equation}
	S= \lim_{\epsilon\rightarrow 0} \frac{d a_d}{d-1}T^{d-1} 
	\frac{\ell^{d-1}}{\epsilon^{d-1}}L^{d-1} ,
	\label{eq4046}
\end{equation}
where $(\ell^{d-1}/\epsilon^{d-1})L^{d-1}$ is the area of the hypersurface at
$\rho=\epsilon$, which approaches the boundary as $\epsilon \rightarrow 0$.
As in the case of AdS Schwarzschild BH, we assume that the total entropy (\ref{eq4046}) of the CFT at the holographic boundary 
 equals the 
AdS planar BH entropy (\ref{eq3046}).
Then, from the equation 
$S=S_{\rm pl}$ we find  
\begin{equation}
	T^{d-1}=\frac{d-1}{4d a_d G_{d+1}}\frac{\epsilon^{d-1}}{\ell^{d-1}}	\frac{\rho_{\rm pl}^{d+1}}{\ell^{d+1}} . 
	\label{eq4047}
\end{equation}
As mentioned in Sec.\ \ref{entropy}, to recover the area law for the planar BH 
we set $\rho_{\rm pl}=\ell$. In this case, according to (\ref{eq1011}),
the temperature of the planar BH is $T_{\rm pl}=d/(4\pi\ell)$. Using this
and 
Eq.\ (\ref{eq4047}), the temperature $T$ may be expressed as 
\begin{equation}
	T=\frac{4\pi \ell}{d}\left( \frac{d-1}{4d a_d G_{d+1}}\right)^{1/(d-1)}    	\frac{T_{\rm pl}}{\ell/\epsilon} . 
	\label{eq4048}
\end{equation}
Here, similar to case of Schwarzschild BH geometry, the quantity $T_{\rm pl}/(\ell/\epsilon)$ can be regarded as  the local temperature  near the AdS boundary at $\rho=\epsilon$,
\begin{equation}
	T_{\rm pl,loc}\equiv\frac{T_{\rm pl}}{\sqrt{g_{00}}} =\frac{T_{\rm pl}}{\ell/\epsilon} .
	\label{eq4049}
\end{equation}
Hence, the expression (\ref{eq4048}) relates the temperature of the holographic conformal fluid to the planar BH  local temperature near the  AdS boundary. 

Now,  we identify the temperature $T$ of the boundary conformal fluid defined
by (\ref{eq4048}) with that defined by (\ref{eq2048}).
As a consequence, we obtain a relation between the redshifted local temperatures of the planar BH and the large Schwarzschild BH measured near the AdS boundary
\begin{equation}
\frac{T_+}{R/\ell} =\frac{T_{\rm pl}}{\ell/\epsilon} .
\label{eq1033}
\end{equation}
	Here, the quantity $R$ is a cutoff near the AdS boundary with  
	spherically symmetric geometry, and $\epsilon$ is the corresponding cutoff
	with planar geometry. We introduce these cutoffs  to regularize the entropy of the conformal fluid at the AdS boundary, so it is
	natural to assume that $R/\ell$ and $\ell/\epsilon$ are of the same order of magnitude.
Hence, the large AdS Schwarzschild BH appears 
as hot as the corresponding planar BH. 
Using (\ref{eq1033}) and (\ref{eq1011}) with $\rho_{\rm pl}=\ell$ we can express 
the horizon temperature of the 
large AdS Schwarzschild BH in terms of the planar BH temperature
\begin{equation}
	T_+ \sim T_{\rm pl}=\frac{d}{4\pi\ell} .
	\label{eq1034}
\end{equation}

Equations (\ref{eq1033}) and (\ref{eq1034}), as our main result,
are obtained 
by invoking the AdS/CFT conjecture and assuming that the temperature of an AdS planar BH can be fixed by
imposing the entropy area law. 
	The main assumption leading to equation (\ref{eq1033}) is the identification of the total entropy of the CFT at the holographic boundary with the 
	horizon entropy of the BH in the bulk. In the case of spherical symmetry,
	this identification yields the black-body radiation temperature of the conformal fluid expressed in terms of the BH horizon temperature $T_+$ via equation (\ref{eq2047}).
	Similarly, in the case of planar symmetry, 	the same identification yields
	the black-body temperature of the conformal fluid expressed in terms
	of the planar BH temperature $T_{\rm pl}$ via equation  (\ref{eq2048}).
	Combining equations  (\ref{eq2047}) and  (\ref{eq2048}) we obtain (\ref{eq1033}). Then, equation (\ref{eq1034}) is a direct consequence of (\ref{eq1033}) since   the cutoffs $R/\ell$ and $\ell/\epsilon$  are assumed to be of the same order of magnitude.
	The physical meaning of equations (\ref{eq1033}) and (\ref{eq1034}) is 
	quite clear: the temperature 
	of a large Schwarzschild BH and that of the corresponding planar BH, as seen by an observer near the AdS boundary, are of the same order of magnitude.
	This result is significant as it resolves the horizon temperature ambiguity of the large AdS Schwarzschild BH and its relation to the AdS planar BH.

%Hence, observers in free fall outside a large AdS black hole
%never see thermal radiation at the Hawking temperature  \cite{hemming,gregory,hubeny}.
\section{Conclusions}
\label{conclude}
By making use of the AdS/CFT correspondence we have calculated the traceless part of 
the vacuum expectation value of the stress tensor of the boundary CFT corresponding to the bulk geometry of AdS Schwarzschild and planar BHs.
Based on the correspondence between the CFT entropy at the boundary and the BH entropies, 
we have established relationships between the temperatures $T$, $T_+$, and $T_{\rm pl}$ of the  conformal fluid at the boundary, 
large Schwarzschild BH, and planar BH, respectively.

In this way, we  predict that the temperatures of large AdS Schwarzschild and planar BHs 
are of the same order of magnitude as seen by an observer near the AdS boundary.
 Because of the diverging  redshift near the boundary,
	local observers  measure a large redshifted temperature of the AdS Schwarzschild BH 
	and the actual horizon temperature $T_+$ is unobservable.
	This contrasts with the case of ordinary
	Schwarzschild BH  where asymptotic observers see the actual horizon temperature. 

Our result confirms the argument of Ref.\ \cite{hubeny} based on 
the translational invariance of the planar BH horizon and previous assertion
\cite{hemming,gregory}
that the high temperatures associated with large values of $T_+$ are not locally observable.
\section*{Acknowledgments}

N.~Bili\'c has been partially supported  by the ICTP - SEENET-MTP project NT-03 Cosmology - Classical and Quantum Challenges. 
J.~C.~Fabris thanks  CNPq (Brazil) and FAPES (Brazil) for partial support.

\end{document}